\documentclass[aps,prd,preprint,superscriptaddress,showpacs,floatfix,nobibnotes,nofootinbib]{revtex4-1}
\usepackage{geometry}
\usepackage{latexsym}
\usepackage{amsmath}
\usepackage{amssymb}
\usepackage{graphicx}
\usepackage{setspace}
\usepackage{slashed}

\usepackage{float}
\usepackage{caption}
\usepackage{subcaption}
\usepackage{bm}
\usepackage{epsfig}
\newcommand{\bea}{\begin{eqnarray}}
\newcommand{\eea}{\end{eqnarray}}

\usepackage[svgnames]{xcolor}

\begin{document}

\title{The effect of anisotropy on phase transitions in graphene}

\author{M.E. Carrington}
\email[]{carrington@brandonu.ca} 
\affiliation{Department of Physics, Brandon University, Brandon, Manitoba, R7A 6A9 Canada}
\affiliation{Department of Physics \& Astronomy, University of Manitoba, Winnipeg, Manitoba, R3T 2N2 Canada}
\affiliation{Winnipeg Institute for Theoretical Physics, Winnipeg, Manitoba}
 
\author{A.R. Frey}
\email[]{a.frey@uwinnipeg.ca} \affiliation{Department of Physics, University of Winnipeg, Winnipeg, Manitoba, R3M 2E9 Canada}
\affiliation{Department of Physics \& Astronomy, University of Manitoba, Winnipeg, Manitoba, R3T 2N2 Canada}
\affiliation{Winnipeg Institute for Theoretical Physics, Winnipeg, Manitoba}

\author{B.A. Meggison}
\email[]{brett.meggison@gmail.com} 
\affiliation{Department of Physics \& Astronomy, University of Manitoba, Winnipeg, Manitoba, R3T 2N2 Canada}
\affiliation{Winnipeg Institute for Theoretical Physics, Winnipeg, Manitoba}

\date{\today}

\begin{abstract}
We study the effect of anisotropy (strain)  on dynamical gap generation in graphene. 
We work with  a low energy effective theory obtained from a tight-binding Hamiltonian expanded around the Dirac points in momentum space.
We use a non-perturbative Schwinger-Dyson approach and calculate a coupled set of five momentum dependent dressing functions. 
Our results show that the critical coupling depends only weakly on the anisotropy parameter, and increases with greater anisotropy. 
\end{abstract}


\normalsize
\maketitle

\normalsize

\section{Introduction}
\label{section-introduction}

There has been tremendous recent interest in the physics of graphene. 
This is in part because of promising graphene-based  technological applications including
transistors, optoelectronics, and many others. 
One significant problem is that due to the lack of finite spectral gap at the charge neutrality point, the material cannot be directly used for certain electronics applications. 
There have been many proposals to generate a mass gap in graphene, or equivalently to induce a transition from the semi-metal state to that of an insulator. A popular proposal that we will focus on in this paper, is to use structural changes (strain), which are known to alter the electronic band structure of graphene \cite{pereira}. 
The effect of strain on graphene is also of practical importance as related to the mechanical strength of the material and its potential use in developing stretchable, transparent, and carbon based nanoelectronics devices.

Graphene is also of fundamental interest to theoretical physicists for a number of reasons. 
Because of its particular lattice structure,
the low energy dynamics are described by a continuum quantum field theory in which the
electronic quasi-particles have a linear Dirac-like dispersion relation of the form $E = \pm v_F p$
where $v_F \sim c/300$ is the velocity of a massless electron in graphene.
The system can be described using reduced quantum electrodynamics (RQED$_{3+1}$), in which the fermions are restricted to move in the two-dimensional plane of the graphene sheet, while the photons are free to move in three dimensions \cite{marino, miransky}. 
The coupling constant in the theory is dimensionless, and the interaction between the electrons has the same $1/r$ Coulomb form as in the (3+1) dimensional theory (and not the $\ln(r)$ dependence of the (2+1) dimensional formulation of QED). 
In addition, renormalization of the theory involves only a single momentum independent subtraction, and is therefore essentially trivial. 
On the other hand, RQED$_{3+1}$ is strongly coupled and in this sense more complicated than QED. The theory therefore plays the role of an interesting toy model to study non-perturbative effects in QCD, which has a much more complicated divergence structure, in addition to being non-abelian.

Anisotropic RQED$_{3+1}$ has been used previously to study graphene by a number of authors. 
Ref.~\cite{wang-anio} used a renormalization group method, working to leading order in $1/N$ where $N$ is the flavour of Dirac fermions. They found that the dynamical gap is suppressed as  anisotropy increases. 
In this paper we will use a Schwinger-Dyson (SD) approach\footnote{For reviews see \cite{roberts-review,roberts-review-2}.}.
Related calculations have been done previously by two groups \cite{sharma1, sharma2} and \cite{xiao1}. 
The results do not do not agree with each other, but the difference might be caused by differences in the way that anisotropy was defined in combination with the approximations that were used.\footnote{The authors of Ref. \cite{xiao1} argue that the effective coupling in \cite{sharma1, sharma2} is not defined in a way that makes it possible to introduce anisotropy without also changing the coupling, which means that the anisotropy and the coupling are not really independent parameters.}
In this paper, we try to clarify this situation by performing a more general calculation in which all fermion dressing functions are determined self-consistently. 
The SD equations for anisotropic graphene involve a large number of non-perturbative dressing functions, because some of the symmetries of the corresponding vacuum field theory are not present. 
The non-relativistic Fermi velocity breaks Lorentz invariance. To study anisotropy we must also break the two-dimensional spatial symmetry. 
Both of these features require the introduction of additional dressing functions, which significantly increase the difficulty of the calculation. 

It is commonly argued that not all of these dressing functions are necessary. The idea is that one can make many simplifying assumptions, and still obtain a qualitative picture of the phase transition. 
The resulting numerical simplifications are significant, and the approach seems particularly reasonable if one only wants to obtain information about whether or not anisotropy enhances or suppresses gap formation. However, since the contradictory results obtained in previous works could well be caused by an artifact of the approximations that were used, it is important to perform a full calculation in which all fermion dressing functions are determined self-consistently.
It is known that, for isotropic graphene, the inclusion of these dressing functions impacts the critical coupling significantly \cite{fischer, mec1, mec2}, which suggests they could also play an important role in the anisotropic system.

A set of integral equations to perform this calculation was derived in \cite{xiao1}; however, there is an internal inconsistency with their formalism. This problem does not affect their numerical calculations, since the problem disappears in the approximation that all dressing functions except the gap function are set to their bare values, but it does mean that the equations they derived are not suitable for the calculation we are going to do. The origin of the problem is easy to describe.
We first note that the Euclidean space inverse propagator for a Lorentz-invariant fermion can be written in terms of two dressing functions as 
$S^{-1}(P) = -i\big(A\slashed{P}+D\big)$, where $(A,D)$ are momentum dependent scalar functions. 
In the isotropic low energy effective theory that describes graphene, 
the non-relativistic Fermi velocity breaks Lorentz invariance, which requires 
a third dressing function. Using the notation of \cite{mec1,mec2}, the inverse 
propagator has the form\footnote{Note 
that we use the `slash' notation in a transparent but somewhat unconventional 
way to denote any quantity contracted with a gamma matrix, even if the result 
is not a Lorentz scalar. For example $\slashed{p}_0 = \gamma_0 p_0$.} 
$-i\big(Z \slashed{p}_0+v_FA\slashed{\vec p} + D\big)$.
In an anisotropic system, where we need a fourth dressing function, we could write the inverse propagator as $-i(Z \slashed{p}_0+v_{1} A_1 \slashed{p}_1+v_{2} A_2 \slashed{p}_2 + D)$, and $v_1\ne v_2$. 
This construction seems natural, since $Z=A_1=A_2=1$, $v_1=v_2=v_F$ and $D=m$ reduces to the bare inverse propagator, and $A_1=A_2=A$, $v_1=v_2=v_F$ reproduces the isotropic expression. 
The results of Refs. \cite{sharma1,sharma2,xiao1} are obtained by setting $Z=A_1=A_2=1$, using bare vertices, and solving a single integral equation for the dressing function $D$. 
One could try to improve this calculation by solving a coupled set of integral equations for the four fermion dressing functions. 
However, setting $v_1=v_2$ and $A_1 = A_2$ does not give a solution of these equations. Furthermore, neither $A_1$ nor $A_2$ satisfies the equation obtained for the dressing function $A$ by taking the appropriate projection of the fermion SD equation  in the isotropic theory. 
We see therefore that when the non-perturbative calculation is formulated in this way, the isotropic limit does not produce the isotropic solution.

In this paper we introduce four fermion dressing functions, as described above, but use a different construction for the non-perturbative fermion propagator, which correctly reduces to the isotropic result in the appropriate limit. 
We calculate all four dressing functions  self-consistently, and keep all frequency dependence. 
We use the common one-loop approximation for the photon polarization tensor, 
which is justified by the vanishing electron density of states at the Dirac points. 
To reduce the numerical problem to a tractable level, we truncate the hierarchy of SD equations using a vertex ansatz which allows us to avoid introducing additional vertex dressing functions. The construction of vertex ans\"atze that preserve gauge invariance and are well adapted for calculational efficiency has been studied in many papers; see for example  \cite{ball-chiu-1,ball-chiu-2,pennington-1,pennington-2,pennington-3}. The vertex ansatz that we use is discussed in section \ref{setup2}. 

It is worth noting that $n$-particle-irreducible ($n$PI) approaches have the advantage, relative to SD methods, that all truncations occur at the level of the action, and gauge invariance is respected to the order of the truncation \cite{Smit2003,Zaraket2004}. In addition, a method has recently been developed to renormalize the effective action, up to the 4PI level \cite{4pi-renorm1,4pi-renorm2,4pi-renorm3}. 
However, these methods are also numerically challenging and have not yet been applied to a four-dimensional gauge theory beyond the leading (2PI) level.  Because of these technical difficulties, we use an SD approach.
The main issue with this method is that one obtains an infinite coupled hierarchy of integral equations for the $n$-point functions of the theory, which must be truncated by introducing an ansatz as described above.

Finally we comment that in any calculation based on an effective theory, there are potentially important screening effects that are necessarily ignored. 
The inclusion of screening from the $\sigma$-band electrons and localized higher energy states
requires a lattice-based approach, but these calculations typically employ the Coulomb approximation and therefore neglect frequency effects. 
For example,
Ref.~\cite{PhysRevLett.115.186602} used a quantum Monte-Carlo simulation on a honeycomb lattice with both
Hubbard and Coulomb interactions between electrons. They found that the short distance screening effects enhance the transition to the insulating 
state.\footnote{We note also that \cite{PhysRevLett.115.186602} considers 
isotropic strain, so it does not address the question of most interest to us.}

The value of the critical coupling produced by a calculation based on either a low energy effective theory, or a honeycomb structured lattice calculation, is not expected to be exact. 
The goal is to explore the influence and relative importance of different physical effects. The point of the calculation done in this paper is to establish whether or not anisotropy could reduce the critical coupling, and therefore make it experimentally possible to produce an insulating state. Our results indicate anisotropy increases the critical coupling, instead of moving it downward toward values that could be physically realizable.

This paper is organized as follows. In section \ref{setup} we define our notation and derive the set of SD equations that we will solve. 
In section \ref{numerics} we describe our numerical method. We present and discuss our results in section \ref{results}, and some conclusions are given in section \ref{conclusions}. We use throughout natural units ($\hbar=c=1$). We work in Euclidean space and use capital letters and Greek indices for (2+1)-dimensional vectors: for example $P_\mu = (p_0,p_1,p_2) = (p_0,\vec p)$ and $P^2=p_0^2+p^2$. For integration variables we use, for example, $dK = \int dk_0\,d^2k/ (2\pi)^3\,.$ We define $Q=K-P$. We frequently abbreviate the arguments of scalar functions, for example $D(P)\equiv D(p_0,\vec p)$.

\section{Physical Set-Up}
\label{setup}

\subsection{Propagators and Dressing Functions}

The Euclidean action of the low energy effective theory is 
\begin{equation}
\label{action}
S=\int d^3 x \sum_{a}\bar\psi_a \left(i\partial_\mu -e A_\mu\right)M_{\mu\nu}\gamma_\nu \psi _a - \frac{\epsilon}{4e^2}\int d^3x F_{\mu\nu}\frac{1}{2\sqrt{-\partial^2}}F_{\mu\nu} \text{ + gauge fixing}.
\end{equation}
The gauge field action is non-local because the photon which mediates the interactions between the electrons propagates in the 3+1 dimensional space-time, and therefore out of the graphene plane.
The fermionic part of the action looks like a free Dirac theory with a linear dispersion relation, because the effective theory describes the system close to the Dirac points. 
We use a representation of the three four-dimensional
$\gamma$-matrices that satisfy $\{\gamma_\mu,\gamma_\nu\} = 2\delta_{\mu\nu}$.
The Feynman rules for the bare theory, in covariant gauge, are
\bea
\label{bareFR}
&& S^{(0)}(P) = -\big[i\gamma_\mu M_{\mu\nu} P_\nu\big]^{-1}\,\\[2mm]
&& G^{(0)}_{\mu\nu}(P)=\big[\delta_{\mu\nu}-\frac{P_\mu P_\nu}{P^2}(1-\xi)\big]\,\frac{1}{2\sqrt{P^2}}\, \\[1mm]
\label{barevert}
&& \Gamma^{(0)}_\mu = M_{\mu\nu}\gamma_\nu\,
\eea
where we have defined
\bea
\label{Mdef}
M = 
\left[\begin{array}{ccc}
~1~ & ~0~ & ~0~ \\
0 & v_{1}  & 0 \\
0 & 0 & v_{2}   \\
\end{array}
\right]\,.
\eea
In the isotropic limit, $v_1=v_2=v_F\equiv c/300$; we call
$v_1$ and $v_2$ the principal velocities (with principal axes in the 1,2
directions). The Fermi velocity is the geometric mean $v_F=\sqrt{v_1v_2}$,
and the anisotropy parameter is the ratio $\eta=v_1/v_2$.

To write the non-perturbative photon propagator, we
define the projection operators 
\bea
&& P^1_{\mu\nu}=\delta_{\mu\nu}-\frac{P_\mu P_\nu}{P^2}\,,~~ P^2_{\mu\nu}=\frac{P_\mu P_\nu}{P^2}\,,~~ P^3_{\mu\nu} = \frac{N_\mu N_\nu}{N^2},
\eea
where $N_\mu=\delta_{\mu 0}-p_0P_\mu/P^2$.
The photon polarization tensor is defined by the equation
\bea
G_{\mu\nu}^{-1} = \frac{2}{\sqrt{P^2}}P^2\big(P_{\mu\nu}^1+\frac{1}{\xi} P^2_{\mu\nu}\big)+\Pi_{\mu\nu}\,.
\eea
Inverting this expression we obtain the dressed propagator, and in 
Landau gauge ($\xi=0$) we have 
\bea
\label{fullG}
&& G_{\mu\nu} =  
\frac{P_{\mu\nu}^1}{G_T(p_0,\vec p)}+ P_{\mu\nu}^3\left(\frac{1}{G_L(p_0,\vec p)}-\frac{1}{G_T(p_0,\vec p)}\right)\, ,\nonumber\\[4mm]
\label{fullPi}
&& G_T(p_0,\vec p) = 2\sqrt{P^2}+\alpha(p_0,p)\, ,\nonumber\\
&& G_L(p_0,\vec p) = 2\sqrt{P^2}+\alpha(p_0,p)+\gamma(p_0,p) \,
\eea
where the dressing functions $\alpha$ and $\gamma$ are related to the trace and zero-zero component of the polarization tensor as
\bea
&& {\rm Tr}\Pi(p_0,p) = \big(2\alpha(p_0,p) + \gamma(p_0,p)\big)\, \nonumber\\
\label{00L}
&& \Pi_{00}(p_0,p) = \frac{p^2}{P^2}\big(\alpha(p_0,p) + \gamma(p_0,p)\big)\,. 
\eea

The fermion self energy is defined through the equation
\bea
\label{SF2}
&& S^{-1}(P) = (S^{(0)})^{-1}(P)+\Sigma(P)\,.
\eea 
The dressed fermion propagator is written in terms of four independent dressing functions which we denote $Z(p_0,\vec p)$, $A_1(p_0,\vec p)$,  $A_2(p_0,\vec p)$ and $D(p_0,\vec p)$. We will sometimes write the arguments as a single subscript so that the dressing functions are denoted $Z_p$, $A_{\rm 1p}$, $ A_{\rm 2p}$ and $D_p$.
We define the matrix
\bea
F(p_0,\vec p) = \left[
\begin{array}{ccc}
 Z_p & 0 & 0 \\
 0 & A_{\text{1p}} & A_{\text{2p}} \\
 0 & -A_{\text{2p}} & A_{\text{1p}} \\
\end{array}
\right]\,\label{F-def}
\eea
and the inverse propagator takes the form
\bea
S^{-1} = -i \gamma_\mu M_{\mu\alpha}F(p_0,\vec p)_{\alpha\nu}P_\nu + D_p\,.
\eea
We note that more general ansaetz are possible, see for example \cite{haldane_88,Carrington_2019}. 
Inverting the inverse propatator we obtain
\bea
S = \frac{1}{S_p}\left[ i \gamma_\mu M_{\mu\alpha} F(p_0,\vec p)_{\alpha\nu} P_\nu +D_p \right]
\eea
with
\bea
S_p = p_0^2 Z_p^2 + v_1^2 \left(p_1 A_{\text{1p}}+p_2 A_{\text{2p}}\right){}^2 
+ v_2^2 \left(p_2 A_{\text{1p}} - p_1 A_{\text{2p}}\right){}^2+D_p^2 \,.\label{Sp}  
\eea
Comparing with equation (\ref{bareFR}) it is clear that the bare theory is obtained by setting $Z(p_0,\vec p) = A_1(p_0,\vec p) =1$ and $A_2(p_0,\vec p)= D(p_0,\vec p)=0$.

The dressing functions $Z$, $A_1$, and $A_2$ when written as the 
matrix $F$ in equation (\ref{F-def}) describe the renormalization of the tensor $M$, i.e., $\hat M=MF$.
To interpret $\hat M$, note that the renormalized (Euclidean) dispersion
relation $S_p=0$ is
\bea
&&\left[\begin{array}{ccc}p_0 & p_1&p_2\end{array}\right]
\hat M^T\hat M
\left[\begin{array}{ccc}p_0 \\ p_1\\ p_2\end{array}\right] + D^2=0
\label{dispE1}
\eea
(where we have suppressed the momentum dependence of the dressing functions). 
Close to the critical point we can set $D=0$ and rewrite (\ref{dispE1}) in the basis formed by the eigenvectors of
\bea
&& \hat M^T\hat M = 
\left(
\begin{array}{ccc}
 Z^2 & 0 & 0 \\
 0 & A_1^2 v_1^2+A_2^2 v_2^2 & A_1 A_2 \left(v_1^2-v_2^2\right) \\
 0 & A_1 A_2 \left(v_1^2-v_2^2\right) & A_2^2 v_1^2+A_1^2 v_2^2 \\
\end{array}
\right)\,.\label{matt}
\eea
In this basis the dispersion relation takes the perturbative form 
\bea p_0^2 +(\hat v_1)^2p_1^2+(\hat v_2)^2p_2^2=0\eea
where the renormalized principal velocities 
$\hat v_1=v_1\sqrt{A_1^2+A_2^2}/Z$ and
$\hat v_2=v_2\sqrt{A_1^2+A_2^2}/Z$ are given by 
(square roots of) the eigenvalues of (\ref{matt}). 
We see that the renormalized Fermi velocity is
$v_F\sqrt{A_1^2+A_2^2}/Z$ and the anisotropy parameter is not renormalized.

\subsection{Schwinger-Dyson equations}
\label{setup2}

The SD equation for the fermion self energy is
\bea
\label{fermion-SD}
&& \Sigma(p_0,\vec p) = e^2\int dK \,G_{\mu\nu}(q_0,\vec q)\,M_{\mu\tau}\,\gamma_\tau \, S(k_0,\vec k) \,\Gamma_\nu\, ,
\eea
and the SD equation for the polarization tensor is
\bea
\label{photon-SD}
&& \Pi_{\mu\nu}(p_0,\vec p) = -e^2\int dK \,{\rm Tr}\,\big[S(q_0,\vec q) \, M_{\mu\tau} \, \gamma_\tau \, S(k_0,\vec k)\, \Gamma_\nu\big]\,.
\eea
To leading order in $(v_1/c,v_2/c)$ the only component of the propagator (\ref{fullG}) that contributes to the fermion self energy $\Sigma$ is the piece $G_L$, so we only need to calculate the zero-zero component of the polarization tensor (see equation (\ref{00L})). 

The three-point vertex in equations (\ref{fermion-SD}, \ref{photon-SD}) should, in principle, be determined from its own SD equation. 
Vertex functions are extremely difficult to calculate numerically, so we introduce an ansatz for the three-point function, which effectively  truncates the hierarchy of SD equations. 
The original Ball-Chiu vertex ansatz \cite{ball-chiu-1,ball-chiu-2} preserves gauge invariance in a Lorentz invariant theory. A modified version of this ansatz that satisfies gauge invariance in our anisotropic theory is
\bea
\Gamma_\mu(P,K) && = \frac{1}{2}\big[F(p_0,\vec p)_{\mu\alpha}^T+F(k_0,\vec k)_{\mu\alpha}^T\big]M_{\alpha\beta}\gamma_\beta  
 \label{ballchiu} \\
&& +\bigg[\frac{1}{2}(P+K)_\alpha\big[F(p_0,\vec p)_{\alpha\beta}^T-F(k_0,\vec k)_{\alpha\beta}^T\big]M_{\beta\rho} \gamma_\rho + i(D_p-D_k)\bigg]\frac{(P+K)_\mu}{P^2-K^2}\, ,\nonumber
\eea
where $(P,K)$ are the momenta of the incoming and outgoing fermions, respectively.
This vertex satisfies the Ward identity
\bea
i Q_\mu \Gamma_\mu(P,K) = S^{-1}(p_0,\vec p) -  S^{-1}(k_0,\vec k)\,.
\eea
In numerical calculations, the terms in the second line in (\ref{ballchiu}) are problematic. The reason is that the range of the integration variable ($K$ in our notation) includes the line defined by the equation $K^2=P^2$, and in the limit $K\to P$  these terms approach $0/0 \to$ constant. 
Fortunately, one can check that the contribution from these terms is very small. This was verified for the isotropic calculation in \cite{mec1}, and 
a check for the anisotropic system is currently in progress and will
appear in future work. We therefore proceed using only the first line in the ansatz (\ref{ballchiu}).

We calculate the SD equations for the fermion dressing functions and the zeroth component of the polarization tensor by taking the appropriate projections of (\ref{fermion-SD}) and (\ref{photon-SD}). 
The results are below:
\bea
\label{Znew}
&& Z(p_0,\vec p) = 1-\frac{2\alpha \pi v_F}{p_0}\int \frac{dK}{Q^2 S_k G_L} \,k_0 q^2 Z_k \left(Z_k+Z_p\right)\, , \\
&& A_1(p_0,\vec p) = 1 + \frac{2\alpha \pi v_F}{p^2}\int  \frac{dK}{Q^2 S_k G_L}
\bigg[ k_0 q_0 Z_k (\vec p\cdot\vec q)  \left(A_{\text{1k}} +A_{\text{1p}}+Z_k+Z_p\right) \nonumber \\
&& ~~~~    +q^2 A_{\text{1k}} \left(Z_k+Z_p\right) (\vec k\cdot\vec p)  
    + k_0 q_0 Z_k \left(A_{\text{2k}} + A_{\text{2p}}\right) (\vec p\times \vec q) 
     - q^2 A_{\text{2k}} \left(Z_k+Z_p\right) (\vec k\times \vec p)  \bigg]\, , \nonumber \\ 
\label{A1new}\\
&& A_2(p_0,\vec p) =  \frac{2\alpha \pi v_F}{p^2}\int \frac{dK}{Q^2 S_k G_L}
\bigg[-k_0 q_0 Z_k  (\vec p \times \vec q)  \left(A_{\text{1k}}+A_{\text{1p}}+Z_k+Z_p\right) \nonumber \\
&& ~~~~  +q^2 A_{\text{1k}} \left(Z_k+Z_p\right) (\vec k\times \vec p)  
   + k_0 q_0 Z_k \left(A_{\text{2k}}+A_{\text{2p}}\right) (\vec p\cdot\vec q)  
   + q^2 A_{\text{2k}} \left(Z_k+Z_p\right)  (\vec k\cdot \vec p)  \bigg]\, , \nonumber \\ 
\label{A2new}   \\
&& \label{Dnew} D(p_0,\vec p) = 2\alpha \pi v_F\int \frac{dK}{Q^2 S_k G_L}   \, q^2 D_k \left(Z_k+Z_p\right)\, ,  \\
&& \Pi_{00}(p_0,p) = -16 \pi v_F \alpha \int \frac{dK}{S_k S_q} \bigg[
\left(Z_k+Z_q\right) \left(D_k D_q-k_0 q_0 Z_k Z_q\right)
+ A_{\text{1k}} A_{\text{2q}} \left(Z_k+Z_q\right) (\vec k\times \vec q)_v \nonumber \\
&&~~~~ + A_{\text{1q}} A_{\text{2k}} \left(Z_k+Z_q\right) (\vec q\times \vec k)_v 
+ A_{\text{1k}} A_{\text{1q}} \left(Z_k+Z_q\right) (\vec k\cdot \vec q)_{v}  
+ A_{\text{2k}} A_{\text{2q}} \left(Z_k+Z_q\right) (\vec k\cdot \vec q)_{v'}  \bigg]\, , \nonumber\\ \label{Lnew}
\eea
where we have used the notation
\bea
&& \alpha = \frac{e^2}{4\pi \epsilon v_F}\, , \\
&& \vec k \cdot \vec p = k_1 p_1 + k_2 p_2\, ,  \\
&& (\vec k \cdot \vec p)_v = v_1^2 k_1 p_1 + v_2^2 k_2 p_2\, ,  \\
&& (\vec k \cdot \vec p)_{v'} = v_2^2 k_1 p_1 + v_1^2 k_2 p_2\, ,  \\
&& \vec k \times \vec p = k_1 p_2 - k_2 p_1\, ,  \\
&& (\vec k \times \vec p)_v = v_1^2 k_1 p_2 - v_2^2 k_2 p_1  \,.
\eea
In the isotropic limit ($v_1 = v_2$), equations (\ref{Znew}, \ref{A1new}, \ref{A2new}, \ref{Dnew}, \ref{Lnew}) reduce to 
\bea
\label{ZeqnI}
&& Z_p = 1-\frac{2\alpha\pi v_F}{p_0 } \int \frac{dK}{Q^2 S_k G_L} \,k_0 q^2 Z_k (Z_p+Z_k) \, ,\\
\label{AeqnI}
&& A_{\rm 1p} = 1 + \frac{2\alpha\pi v_F}{p^2} \int \frac{dK}{Q^2 S_k G_L} \,\big[ q^2 A_k (Z_p+Z_k) \vec k \cdot \vec p  +  k_0 q_0 Z_k(Z_p+Z_k + A_p+A_k) \vec p \cdot \vec q \big] \, ,\\
&& A_{\rm 2p} = \frac{2\alpha\pi v_F}{p^2} \int \frac{dK}{Q^2 S_k G_L} \,\big[ q^2 A_{\text{1k}} \left(Z_k+Z_p\right) (\vec k \times \vec p) -k_0 q_0 Z_k (\vec p \times \vec q)  \left(A_{\text{1k}}+A_{\text{1p}}+Z_k+Z_p\right) \big]\, ,\nonumber \\ \\
\label{DeqnI}
&&D_p =  2\alpha\pi v_F \int \frac{dK}{Q^2 S_k G_L} \, q^2 D _k (Z_p+Z_k)\, ,\\[4mm]
\label{LeqnI}
&& \Pi_{00}(p_0,p) = -16 \pi v_F \alpha \int \frac{dK}{S_k S_q} \,\left(Z_k+Z_q\right) \left(A_k A_q v_F^2 (\vec k \cdot \vec q)+D_k D_q-k_0 q_0 Z_k
   Z_q\right)\,.
\eea
The equations for $Z$, $A_1$, $D$ and $\Pi_{00}$ agree with the isotropic calculation of Ref. \cite{mec2}, and it is straightforward to show that $A_2=0$ after performing the integrations. 

We will solve the coupled set of integral equations for the fermion dressing functions (\ref{Znew} - \ref{Dnew}), but we adopt a commonly used approximation, motivated by the vanishing fermion density of states at the Dirac points, which is to use a one-loop result for the polarization component $\Pi_{00}$. Using bare fermion propagators, equation (\ref{Lnew}) gives
\bea
\label{Piana}
\Pi^{\rm 1\,loop}_{00}(p_0,p) = \frac{\pi\alpha}{\sqrt{v_{1} v_{2}}}
\,\frac{p_1^2 v_{1}^2 + p_2^2 v_{2}^2}
{\sqrt{p_0^2+p_1^2 v_{1}^2 + p_2^2 v_{2}^2}} \,.
\eea

We look for solutions to the SD equations with specific symmetry properties which are consistent with the symmetries of the bare theory. The dressing functions $Z$, $A_1$, and $D$ are assumed even under the transformations $p_0\to -p_0$, $p_1\to -p_1$, and $p_2\to -p_2$, and even under the interchange $(p_1,v_1) \leftrightarrow (p_2,v_2)$.
The function $A_2$ is even under $p_0\to -p_0$ and odd under all the other transformations above.
If we assume that these conditions hold under the integrals on the right side of the SD equations, one can show by shifting integration variables that they also hold on the left side; this means that the symmetry conditions we have chosen are satisfied consistently by the equations we solve. 
The interchange $v_1\leftrightarrow v_2$ is equivalent to $\eta \to 1/\eta$ and therefore we expect that the condensate $D(0,0)$ and therefore the critical coupling are invariant under $\eta \to 1/\eta$. We have checked numerically that this condition is satisfied.

\section{Numerics}
\label{numerics}

We use spherical coordinates, so the external momentum variable is represented as $(p_0,p,\theta_p)$, the integration variables are $(k_0,k,\theta_k)$, and 
\bea
dK = \frac{d^3 k}{(2\pi)^3} = \int_{-\infty}^\infty \frac{dk_0}{2\pi} \,\int_0^\infty \frac{dk}{2\pi}\,k\,\int\frac{d\theta_k}{2\pi}. 
\eea
The integration regions for the $k_0$ and $k$ integrals are infinite, but numerically we must use finite bounds. 
This is justified if the theory is properly renormalized, in which case all integrals are ultra-violet finite. 
The only divergence occurs in the photon polarization tensor, and can be removed by a simple subtraction. 
We define
$
\Pi^R_{\mu\nu}(P) = \Pi_{\mu\nu}(P)-\Pi_{\mu\nu}(0)
$,
which satisfies the renormalization condition $\Pi^R_{\mu\nu}(0)=0$.
We perform this renormalization in all numerical calculations and suppress the superscript $R$.
We use a cutoff $\Lambda$ on the momentum integrals. We rescale momenta by $\Lambda$ and dimensionful dressing functions by the appropriate power of $\Lambda$ to remove all dependence on the cutoff.

We use a logarithmic scale for momentum variables to increase the number of grid points close to the origin, where the dressing functions vary the most. In addition, we use Gauss-Legendre quadrature, further increasing the point density around the origin and increasing the overall accuracy of the integration procedure compared to a constant partitioning.

We solve the set of self-consistent integral equations in (\ref{Znew} - \ref{Lnew}) using an iterative procedure. 
The integrands  depend on the dressing functions evaluated at values of $Q=K-P$, which means that interpolation is required. 
After experimentation with several different methods, we determined that the best method for our set of equations is three-dimensional  linear interpolation. 
We have $\vec{q}=\vec{k}-\vec p$, and therefore
\bea
	|q| &=&\sqrt{q_1^2+q_2^2} = \sqrt{(k_1-p_1)^2+(k_2-p_2)^2} \nonumber \\
	&=& \sqrt{(k \cos{\theta_k} - p \cos{\theta_p})^2 + (k \sin{\theta_k} - p \sin{\theta_p})^2}\,. 
\eea 
The angle $\theta_q$ is defined through the equation
\bea
\vec q =(q \cos{\theta_q},q \sin{\theta_q})
\eea
and related to the values of $\theta_p$ and $\theta_k$ using a straightforward trigonometric relation
\bea 
\label{q}
\theta_q ={\rm arccos}\left(\frac{k \cos{\theta_{k}} - p\cos{\theta_{p}}}{|\vec k - \vec p|}\right).
\eea

Finally, the integrals that give the fermion dressing functions are numerically unstable because there is a singularity in the integrands when the integration variables $K$ are equal to the external variables $P$. 
This problem is not related to  the anisotropy and appears also in the isotropic calculation. 
It is caused by the factor $1/Q^2 = 1/(K-P)^2$ in the equations for the fermion dressing functions (see equations (\ref{Znew}-\ref{Lnew}) and (\ref{ZeqnI}-\ref{LeqnI})).
These singularities are  integrable, but they must be dealt with carefully in a numerical calculation. 
For example, the $k_0$ integral can be divided into two pieces
$\int_0^\Lambda dk_0 = \int_0^{p_0} dk_0 + \int_{p_0}^\Lambda dk_0$, and 
since Gauss-Legendre is an open integration method that does not use grid points at the exact values of the ends of the integration range, the singular point $p_0$ is not calculated and there is no divergent contribution to the numerical integral. 
In order to obtain a numerically accurate result, the total number of grid points is divided between the two pieces so that the distances between the singularity and the closest points on either side are the same. 

\section{Results}
\label{results}

Our formalism is symmetric under the transformation $\eta\to 1/\eta$, and we have checked that this symmetry is satisfied by the numerical solutions.

\begin{figure}[H]
\centering	
	\begin{subfigure}[h]{0.5\textwidth}
	\centering
	\includegraphics[width=\textwidth]{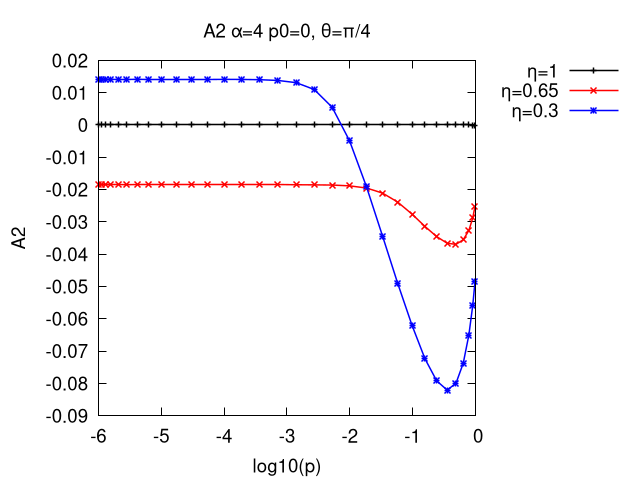}
		\caption{}
	\label{6a}
	\end{subfigure}%
	\hfill
	\begin{subfigure}[h]{0.5\textwidth}
	\includegraphics[width=\linewidth]{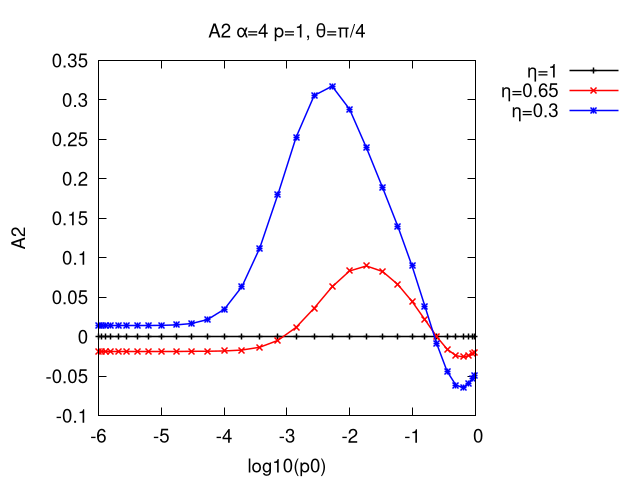}
		\caption{}
	\label{6b}
	\end{subfigure}%
	\caption{$A_2$ dressing function at different $\eta$ showing that $A_2 \rightarrow 0$ as $\eta \rightarrow 1$.}
	\label{6}
\end{figure}

Our equations reduce to the isotropic ones when $\eta=1$, which means that at $\eta=1$ we should find that $A_2$ is zero. This gives a way to test the numerical accuracy of our calculation. In Fig.~\ref{6} we show $A_2(p_0,0)$ for three values of $\eta$; it is visually clear that $A_2$ is comparatively small for $\eta=1$. To obtain a quantitative measure of the size $A_2$ in the isotropic limit, we can integrate over the three dimensional phase space. We find that the ratio $\int d^3 p |A_2^{\eta=1}(p_0,\vec p)| /\int d^3p |A_2^{\eta=.65}(p_0,\vec p)| < 6\times 10^{-4}$.
We have also checked that in the isotropic limit we reproduce the result for the critical coupling obtained in Ref. \cite{mec1}. 

\begin{figure}[H]
\centering	
	\begin{subfigure}[h]{0.5\textwidth}
	\centering
	\includegraphics[width=\textwidth]{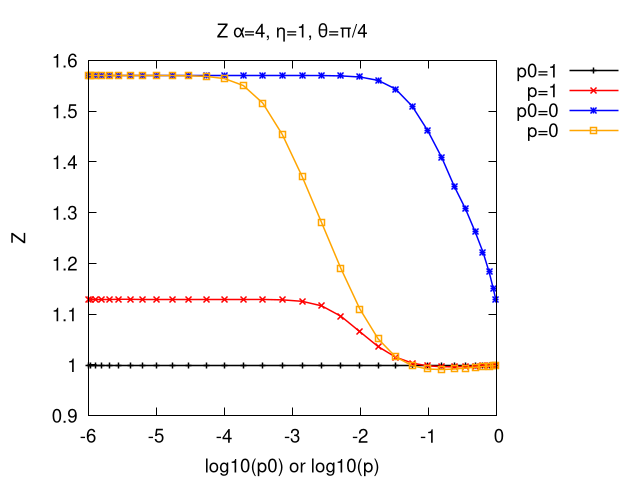}
		\caption{}
	\label{1a}
	\end{subfigure}%
	\hfill
	\begin{subfigure}[h]{0.5\textwidth}
	\includegraphics[width=\linewidth]{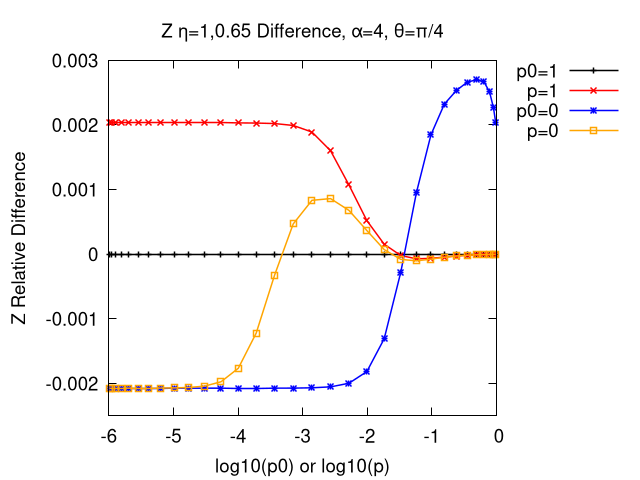}
		\caption{}
		\label{1b}
	\end{subfigure}%
	\caption{$Z$ dressing function for  different cross-sections of  momentum phase space.}
		\label{1}
\end{figure}

\begin{figure}[H]
\centering	
	\begin{subfigure}[h]{0.5\textwidth}
	\centering
	\includegraphics[width=\textwidth]{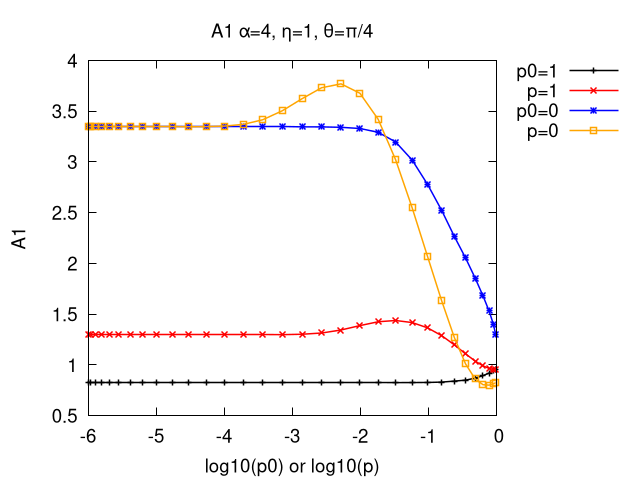}
		\caption{}
	\label{2a}
	\end{subfigure}%
	\hfill
	\begin{subfigure}[h]{0.5\textwidth}
	\includegraphics[width=\linewidth]{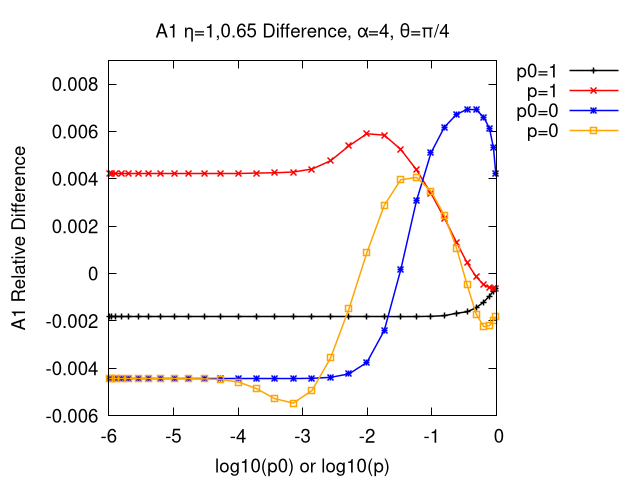}
		\caption{}
	\label{2b}
	\end{subfigure}%
	\caption{$A_1$ dressing function for  different cross-sections of  momentum phase space.}
	\label{2}
\end{figure}

\begin{figure}[h]
\centering	
	\begin{subfigure}[h]{0.5\textwidth}
	\centering
	\includegraphics[width=\textwidth]{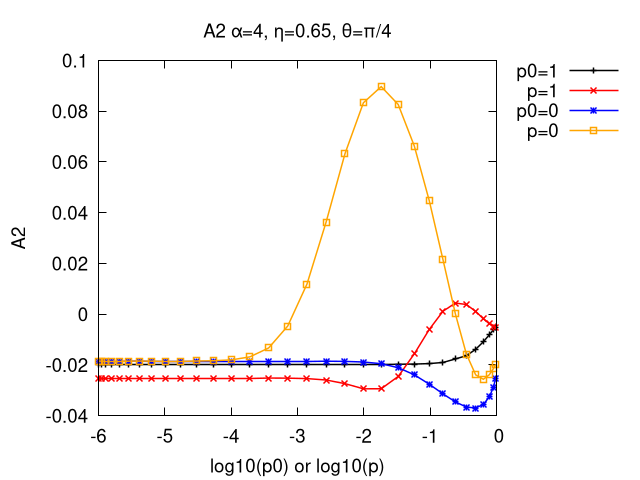}
		\caption{}
	\label{3a}
	\end{subfigure}%
	\hfill
	\begin{subfigure}[h]{0.5\textwidth}
	\includegraphics[width=\linewidth]{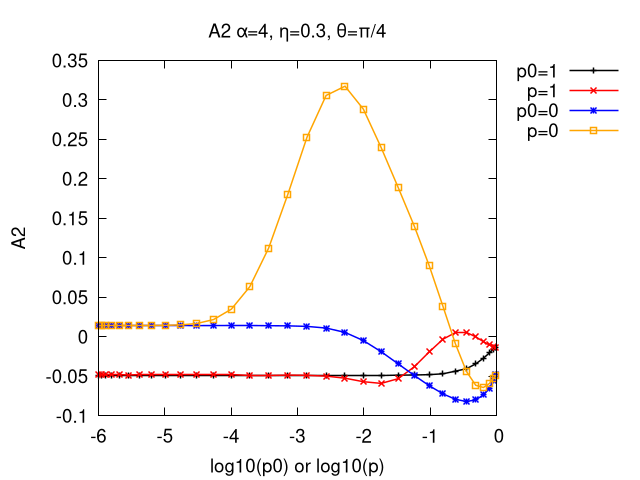}
		\caption{}
	\label{3b}
	\end{subfigure}%
\caption{$A_2$ dressing function for  different cross-sections of  momentum phase space.}
\label{3}
\end{figure}

\begin{figure}[H]
\centering	
	\begin{subfigure}[h]{0.5\textwidth}
	\centering
	\includegraphics[width=\textwidth]{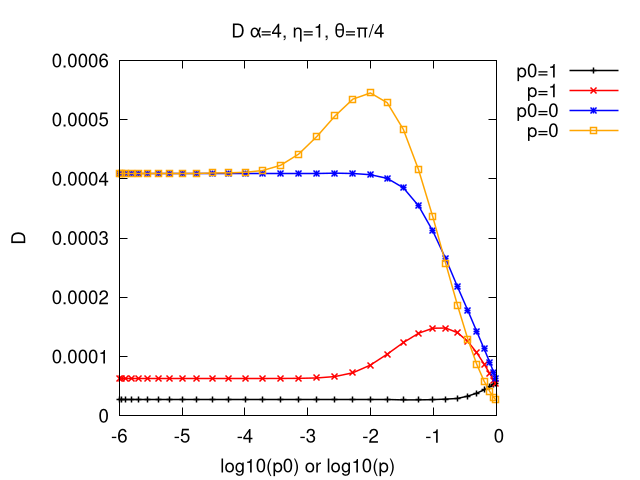}
		\caption{}
	\label{4a}
	\end{subfigure}%
	\hfill
	\begin{subfigure}[h]{0.5\textwidth}
	\includegraphics[width=\linewidth]{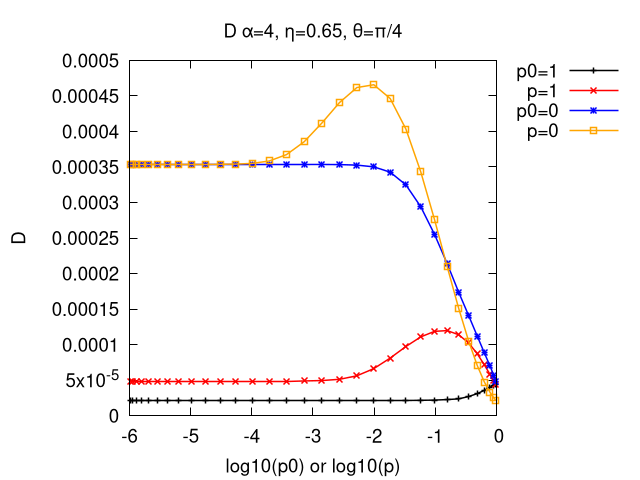}
		\caption{}
	\label{4b}
	\end{subfigure}%
	\caption{$D$ dressing function for  different cross-sections of  momentum phase space.}
	\label{4}
\end{figure}

In Figs. \ref{1}-\ref{4} we show the fermion dressing functions.
The value of the coupling that is shown is $\alpha=4$, which is slightly greater than the critical coupling. 
The value of the angle shown is $\theta=\pi/4$. 
Each graph has four curves, which are obtained by holding either $p_0$ or $p$ fixed, at either its maximum or minimum value (we remind the reader that, using our scaled variables, the maximum value of any momentum variable is 1). 
Figs.~\ref{1a} and \ref{2a} show the isotropic results for the dressing functions $Z$ and $A_1$. The change produced when $\eta$ is reduced from 1 to 0.65 is too small to see on the graph, and therefore Fig.~\ref{1b} shows the relative difference $(Z_{\eta=1}-Z_{\eta=0.65})/(Z_{\eta=1}+Z_{\eta=0.65})$, and Fig.~\ref{2b} shows the same relative difference for $A_1$. 
The dressing function $A_2$ is zero when $\eta=1$, and therefore we show in 
Fig.~\ref{3} two different values of the anisotropy parameter: $\eta=0.65$ and $\eta=0.3$. 
Fig.~\ref{4} shows $D$ for $\eta=1.0$ and $\eta=0.65$. 

\begin{figure}[H]
\centering	
	\begin{subfigure}[h]{0.5\textwidth}
	\centering
	\includegraphics[width=\textwidth]{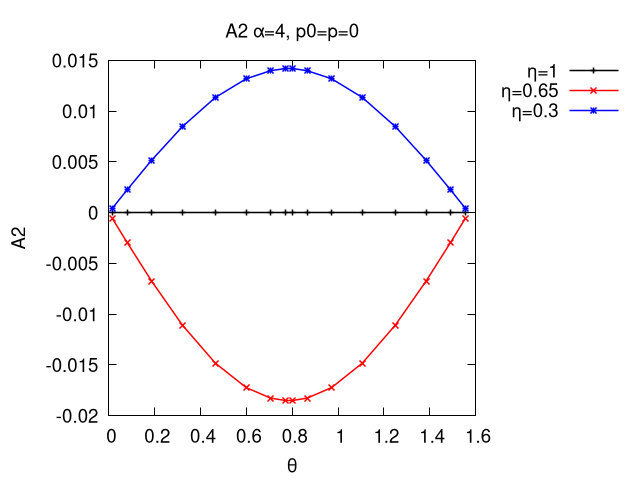}
		\caption{}
	\label{5a}
	\end{subfigure}%
	\hfill
	\begin{subfigure}[h]{0.5\textwidth}
	\includegraphics[width=\linewidth]{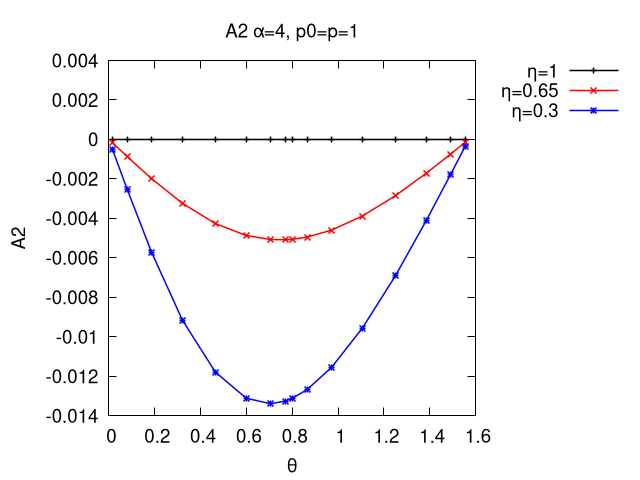}
		\caption{}
	\label{5b}
	\end{subfigure}%
\caption{$A_2$ dressing function vs angle between $p_1$ and $p_2$}
\label{5}
\end{figure}

All dressing functions except $A_2$ depend very weakly on the angle $\theta_p$. 
Fig.~\ref{5} shows the anglular dependence of $A_2$ at large and small momentum, for three different values of the anisotropy parameter. 

We note the following features of these results.
\begin{itemize}
\item At high momentum, all dressing functions approach the perturbative limit ($Z$ and $A_1$ approach 1, while $D$ and $A_2$ approach zero). This verifies that we recover the perturbative limit at high momentum. 

\item The dressing function $A_2$ changes sign close to the zero momentum point when $\eta$ decreases from 0.65 to 0.3, as can be seen by comparing the blue and yellow curves at the left sides of figures \ref{3a} and \ref{3b}. 
We note that the sign change occurs only for small values of both $p_0$ and $p$. 
For both values of $\eta$, the largest contribution occurs at small $p$ and intermediate $p_0$ (the large bumps in the yellow lines in Fig. \ref{3}), and the peak rises and broadens as the anisotropy increases.

\item At low momenta the values of $Z$ and $A_1$ are significantly enhanced (especially $A_1$), which shows the importance of a calculation where all dressing functions are determined self-consistently.
As the coupling is reduced towards the critical coupling, this enhancement becomes even more pronounced. 

\end{itemize}
\begin{figure}[H]
\centering	
	\begin{subfigure}[h]{0.5\textwidth}
	\centering
	\includegraphics[width=\textwidth]{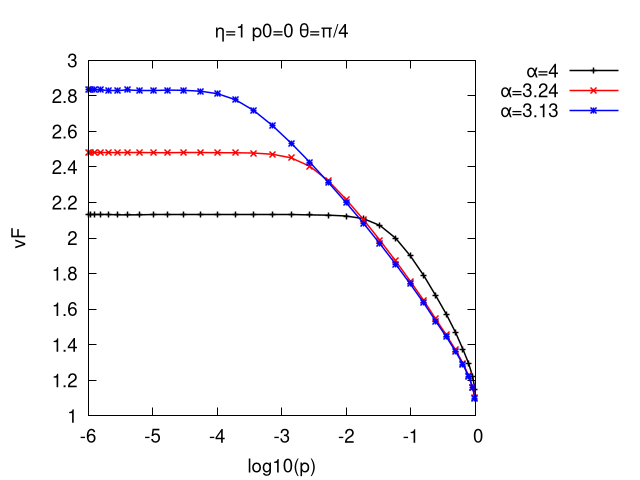}
		\caption{}
	\label{vF}
	\end{subfigure}%
	\hfill
	\begin{subfigure}[h]{0.5\textwidth}
	\includegraphics[width=\linewidth]{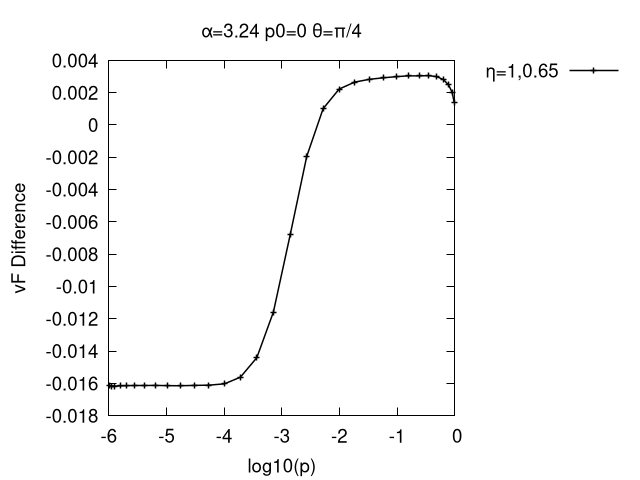}
		\caption{}
	\label{vFdiff}
	\end{subfigure}%
	\caption{The renormalized Fermi velocity.}
\end{figure}

In Fig.~\ref{vF} we show the renormalized Fermi velocity, defined as 
$
v_F = \sqrt{A_1^2 + A_2^2}/Z
$,
versus $p$ with $p_0=0$. Fig.~\ref{vF} shows $\eta = 1$ at $\alpha = 4.0$, $3.24$, and $3.13$. The experimentally observed increase in the Fermi velocity at small coupling \cite{Elias_2011} is clearly seen. Fig.~\ref{vFdiff} shows the difference between $v_F$ at $\eta=1$ and $\eta=0.65$, for $\alpha=3.24$.  As the anisotropy increases, the value of $A_2$ increases, which causes a corresponding increase in the Fermi velocity. 

\begin{figure}[H]
\begin{center}
\includegraphics[width=0.75\linewidth]{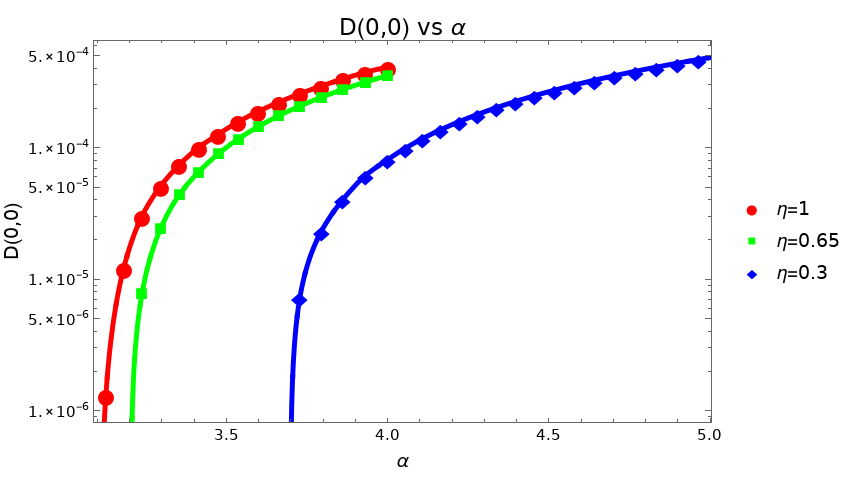}
\end{center}
\caption{The condensate $D(0,0)$ vs. coupling for different values of the anisotropy parameter}
\label{condensate}
\end{figure}

In Fig.~\ref{condensate} we show the value of the condensate $D(0,0)$ versus coupling for three different values of the anisotropy parameter. 
We calculate the critical coupling for the three different values of $\eta$ using the following procedure. 
We consider the inverted function of the data presented in Fig.~\ref{condensate}, i.e. $\alpha[D(0,0)]$ and fit it to a curve.
We then evaluate this function at the value of $\alpha[0]$.
We compare the results obtained from a polynomial fit using polynomials of degree 3 to 5, a Hermite polynomial fit working to orders 3 to 5, and a cubic spline fit. 
The differences between any two fits is less than the quoted uncertainty by at least a factor of 5, which shows that our method for performing the extrapolation does not introduce any appreciable error. 
To obtain a realistic estimate of the uncertainty in our result for the critical coupling, we calculate the difference between the extrapolated result, and 
the result obtained using the same procedure but removing the smallest calculated point.

Our results are shown in the first column of Table \ref{criticalcoupling}. The second column shows the isotropic result obtained using a similar method in Ref. \cite{mec1}. The third column shows the results of \cite{xiao1}, taking into account that the definition of $\eta$ in that paper is equivalent to $1/\eta$ in ours. The numbers quoted  are estimated from their Fig. 7 and are only approximate. The fourth column is the isotropic result from Ref. \cite{fischer} which is obtained using the same approximations as in \cite{xiao1}.
\begin{table}[t]
\caption{Results for critical values of the coupling $\alpha$}
\centering 
\begin{tabular}{|c | c | c|c|c|c|} 
\hline\hline                        
~~ $\eta$ ~~& $~~~~~~~\alpha_c~~~~~~~~$ & ~~~ $\alpha_c$ \cite{mec1} ~~~ & ~~~ $\alpha_c$ \cite{xiao1} ~~~& ~~~ $\alpha_c$ \cite{fischer} ~~~ ~~~ \\ [0.5ex]
\hline                  
1 & 3.12 $\pm$  0.02 & 3.12 $\pm$ 0.01 & $\approx$ 0.92 & 0.09 \\
0.65 & 3.21 $\pm$ 0.02 &  & $\approx 0.94$ & \\
0.3 & 3.70 $\pm$ 0.04 &  & $\approx 1.05$ & \\[1ex]      
\hline
\end{tabular}
\label{criticalcoupling}
\end{table}
The results in Table \ref{criticalcoupling} show that the introduction of anisotropy increases the critical coupling. 
This is consistent with what is seen in Fig.~\ref{vFdiff}, where it is shown that the renormalized fermi-velocity increases as anisotropy increases. This effect supresses the gap, and increases the critical coupling. 
When the fermion dressing functions $Z$ and $A$ are fixed at their perturbative values, as in Refs. \cite{xiao1, fischer}, the effect is missing and the critical coupling that is obtained is greatly reduced.

We comment that the number of iterations required to converge to a solution of the SD equations increases significantly as $\alpha$ approaches the critical value, due to what is known as `critical slowing down.' 
This refers generally to a lengthening of the time it takes a system to respond to disturbances when it is close to a critical point (see Ref. \cite{Goldenfeld:1992qy}, section 4.6, for a brief discussion regarding dynamics). 
Mathematically it is easy to see how this problem manifests in our calculation. 
From equation (\ref{Dnew}) it is clear that $D=0$ is always a solution. 
Close to the critical point, the solution we are looking for is very close to this trivial solution, which delays convergence.
When the anisotropy of the system increases, the effect is amplified as the dressing function $A_2$ becomes more important. 
The smallest values of $\alpha$ for which we have obtained solutions require about 600 iterations to converge. 

\section{Conclusions}
\label{conclusions}

We have calculated the critical coupling at which the semi-metal to insulator transition occurs in graphene using a low energy effective theory. 
We have studied the effect of anisotropy on the phase transition, which could be introduced as physical strain on the graphene lattice, or possibly through an applied magnetic field. 
We have included anisotropy by considering a Fermi velocity which is not isotropic in space.
There are several previous calculations in the literature that are similar in their approach \cite{sharma1, sharma2, xiao1}  
but used  numerous restrictive assumptions to make the numerical implementation more tractable. 
The effect of these approximations is difficult to predict, and in fact different approximations have led to predictions that the critical coupling in an anisotropic system moves in different directions, relative to the isotropic one. 
Our calculation includes the complete non-perturbative fermion propagator and a 1-loop photon polarization tensor. Our hierarchy of SD equations are truncated using a Ball-Chiu-like vertex ansatz. Full frequency dependence of the dressing functions is included. 
Our results show that the effect of anisotropy is greater than predicted
by previous calculations, and that it  increases the critical coupling. 

Finally, we remind the reader that the value of the critical coupling produced by any calculation based on an effective theory is not expected to be exact, since there are potentially important screening effects that are necessarily ignored. The point of the calculation is to establish whether or not anisotropy could reduce the critical coupling, and therefore make it experimentally possible to produce an insulating state. Our results indicate anisotropy increases the critical coupling, instead of moving it downward toward values that could be physically realizable. The only significant approximation in our calculation is the use of the 1-loop photon polarization tensor. The back-coupled calculation, in which the polarization tensor is calculated self-consistently together with the fermion dressing functions using equation (\ref{Lnew}) is much more difficult numerically. This calculation is currently in progress.

\begin{acknowledgments}
This work has been supported by the Natural Sciences and
Engineering Research Council of Canada Discovery Grant program.
This research was enabled in part by support provided by WestGrid
(www.westgrid.ca) and Compute Canada Calcul Canada (www.computecanada.ca).
\end{acknowledgments}

\appendix

\section{Numerical convergence}
Our calculation involves solving one loop integral equations in three dimensions. In spherical coordinates, we have three external variables and three integration variables. The numerical calculation therefore involves 6 nested loops. 
The dressing functions themselves are fairly smooth, which means that the number of grid points for the external variables does not have to be very large. However, the integrals involve integrable singularities, which necessitates a  larger number of grid points for the discretized integration variables. Our results were produced using $(N_{p_0}=32)\times (N_p=32) \times (N_{\theta_p}=16)=1.64\times 10^{4}$ external grid points. 
Using the same number of internal grid points, the iteration procedure does not converge to a self-consistent solution. 
We used $(N_{k_0}=100) \times (N_k=100) \times (N_{\theta_k}=32)=3.2\times 10^{5}$ internal grid points, and tested that results are very stable when the number of external and/or internal grid points is increased. 
The total phase space of our calculation contained $\sim  5.2\times 10^9$ grid points. 
We achieved sufficient numerical speed by parallelizing using openMPI 4.0.1.

The number of interations that is needed to achieve convergence increases as the critical point is approached. Convergence can be achieved more quickly, for a given coupling, if the iteration procedure is initialized from the converged data obtained from a numerically similar value of the coupling that has already been calculated.

\bibliography{anisotropic}

\end{document}